\documentstyle[epsf]{lamuphys}
\makeatletter
\let\chapter\hid@chapter
\makeatother
\begin{document}
\pagenumbering{arabic}
\title{An Interpretation of Radio-loud -- Radio-quiet QSO Differences}

\author{Beverley J.\,Wills}

\institute{Department of Astronomy, University of Texas at Austin, Texas,
78712}

\maketitle

\begin{abstract}

Here we speculate on what observations are telling us about the difference
between
radio-loud and radio-quiet QSOs.  The observations are (i) the relation between
ultraviolet-optical luminosity and `jet power', (ii) the dependences of
emission and
absorption line spectra, and the spectral energy distribution, on radio
core-dominance,
assumed to be an indicator of orientation, (iii) the spectral differences
between
radio-loud and radio-quiet QSOs, and (iv) the inverse relation between the
strength of broad, blended Fe\,II multiplets and [O\,III]\,$\lambda$5007, and
the
apparently-related association between Fe\,II strength, reddening, broad
absorption lines,
and scattering polarization.  We present and discuss a picture in which there
are two
main variables: (i) the inclination of the plane of the host galaxy to the axis
of the
inner jet (the central engine's rotation axis), and (ii) the angle of the
line-of-sight
to this rotation axis.  The radio-loud QSOs are those with jets aiming away
from the
plane of the host galaxy.

\end{abstract}
\section{Introduction}

Some hypotheses proposed to explain why $\sim$90\% of QSOs\footnote
{`QSO' refers to all luminous AGN (L$\ga 10^{11}$\,L$_{\sun}$, H$_0 =
 100$\, km s$^{-1}$\,Mpc$^{-1}$.  A radio-loud QSO is one having
F$_{\rm5GHz}$/F$_{4400} \ga 10$, where F is the rest frame flux-density in mJy.
Such strong radio emission is assumed to indicate powerful radio jets.}
 are radio quiet include
(i) an evolutionary phenomenon where radio-loudness is a short-lived phase in
the
existence of all QSOs, or a series of short-lived phases (Schmidt 1970),
(ii) the result of differences in mass concentration in the host galaxy nucleus
(Heckman
1983),
(iii) the result of fundamental angular momentum differences (Wilson \& Colbert
1995),
(iv) the result of poorly-collimated sub-relativistic wind in radio-quiet QSOs
(RQQs)
 (Boroson, Persson \& Oke 1985).

Some hypotheses simply discuss conditions under which jets might form, but
do not attempt to explain all known differences between radio-loud QSOs (RLQs)
and
RQQs.  We take a different approach, by first examining
the relation between ultraviolet-optical luminosity and jet power (see the
chapter,
{\it Accretion and Jet Power}).
There, we concluded that jet power (represented approximately by unbeamed radio
power) is
directly related to the Big Blue Bump luminosity, for RLQs and RQQs,
while the radio luminosity is a factor of $\sim$1000 less in the RQQs.
Then we argued that the generally great similarity of the Big Blue Bump,
non-synchrotron X-ray emission, and the emission line spectra implied a very
similar
central engine mechanism, independent of radio-emission.  This, together with
the
relations between unbeamed radio and Big Blue Bump luminosity, led us to the
hypothesis
that the central engines of RQQs and RLQs (fueling, accretion, and power
available to
generate a jet) are essentially identical.
There is some theoretical support for this, but no single hypothesis is clearly
favoured.

\section{Direct Interpretation of Observations}

Apart from the radio emission, observed differences in photon energy
distribution,
emission-line spectra and absorption spectra may lead to clues concerning the
collimation and propagation of luminous radio jets.  In the previous chapter
{\it Jets and QSO Spectra} we summarized the most significant
relations between radio emission and these ultraviolet--optical properties,
and suggested a consistent picture for the nuclear gas.

For RLQs, we noted that there are dependences of emission-line profiles and
line
strengths on the jet inclination, implying that emission regions are symmetric
about the jet (rotation) axis --- in particular BLR gas velocities are larger
perpendicular
to the jet.  Dependences of profile asymmetry on inclination imply axisymmetric
obscuration
and an axisymmetric velocity field.  Thus we suggested that dust shielded from
the central
ionizing continuum by BLR gas is responsible for revealing high-ionization
axial (polar)
outflow in core-dominant RLQs, and low-ionization equatorial outflow in
lobe-dominant
RLQs.  This equatorial outflow ties in with associated absorption outflows
and increased reddening in lobe-dominant RLQs.
The low-latitude reddening is probably associated with hot AGN dust
as well as the interstellar medium of the host galaxy.  The inner edge of a
dusty torus,
at the evaporation radius for iron-rich grains, may shield from our view Fe\,II
rich gas
produced there, explaining why Fe\,II blends appear weaker in lobe-dominant
RLQs.

These interpretations for RLQs conjure up the dusty-torus model for RQQs
 proposed by Weymann et al.
(1991), in which BAL clouds are ablated from the surface of the torus and
accelerated,
by thermal wind and radiation pressure.  BAL QSOs are just those where the
high-velocity
outflow lies along the line-of-sight.  However, we have little evidence, so
far, for
axisymmetry in RQQs.

Further observational differences between RLQs and RQQs are seen in the inverse
relation
between the strengths of
[O\,III]\,$\lambda$5007 emission from NLR gas at many pcs to Kpcs from the
center, and
the high-velocity Fe\,II emission from the BLR.  RLQs lie at the strong
[O\,III] -- weak
Fe\,II end of this relation, and the BAL QSOs at the weak [O\,III]-- strong
Fe\,II end.  In
this sense the BAL QSOs represent `extreme RQQs'.  We favour an explanation
for this relation in terms of relative covering by dense, high-speed,
Fe\,II-rich gas.  The
lower the covering by dusty low-ionization BLR gas, the more photons are able
to escape
to ionize the more-distant, low-density, NLR gas.

\begin{figure}
\plotone{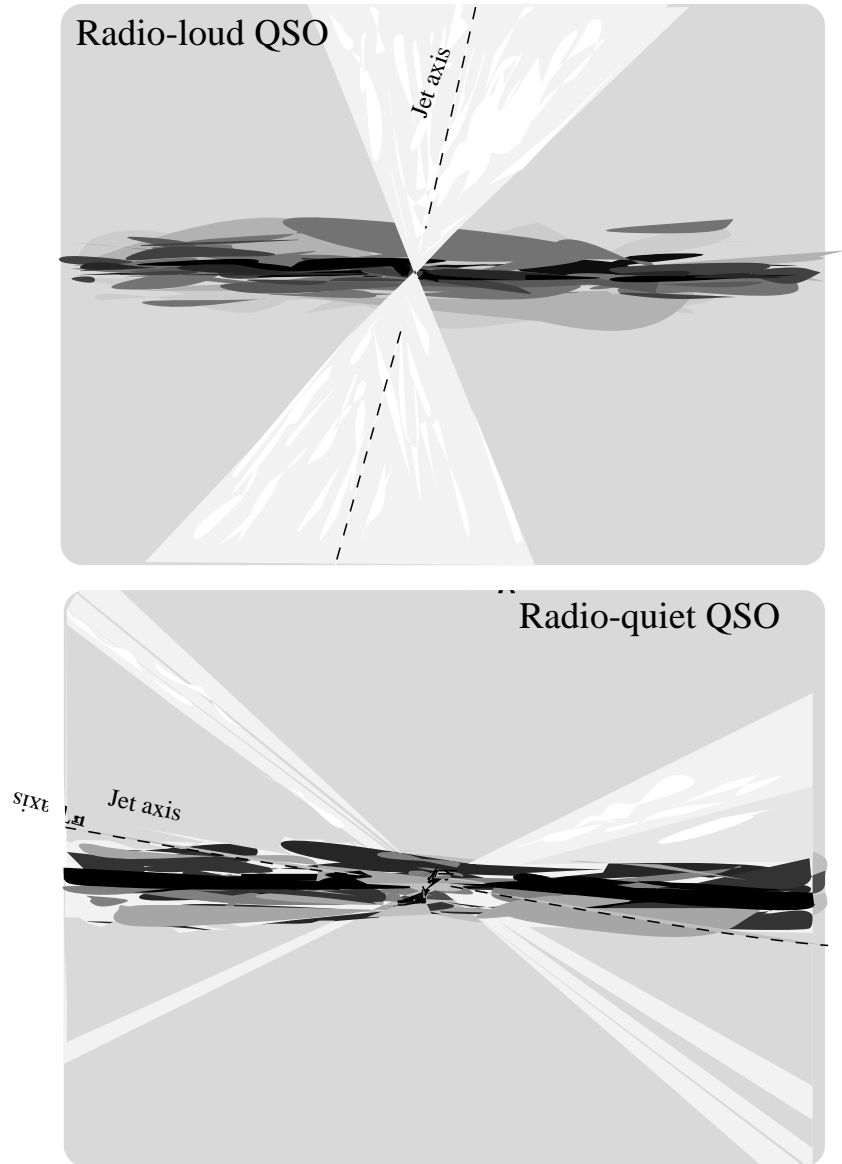}
\caption{
A hypothesis for the different ultraviolet-optical properties of radio-loud and
radio-quiet
QSOs.  The RLQs have jets, and ionization cones formed by the shadow of a dusty
torus,
in a direction away from the highest
concentrations of low-ionization gas and dust in a galaxy, whereas the opposite
is true for
radio-quiet QSOs.  In a planar, axisymmetric, distribution of broad emission
line (BLR) gas,
RLQs' ionizing photons excite gas out of the plane, to distances of several
Kpc.
In radio-quiet QSOs, the NLR gas is partly shielded from ionizing photons by
high-optical-depth broad line gas (Fe\,II emitting), that is ablated from
heated grains near
evaporation temperature.  We show a warped torus whose inner regions are
perpendicular to the
radio axis.
}
\end{figure}

\section{A Possible Model}

The ultraviolet--optical differences between RQQs and RLQs can be
summarized as follows.  Compared with RLQs, RQQs appear to be associated with
more
emission from low-ionization gas (Fe\,II) and show evidence for more
line-of-sight
reddening and high-velocity outflowing (BAL) gas.
The presence of a hot, dusty,
environment with low- and high-ionization (BAL) outflows along the
line-of-sight
therefore has something to do with {\em lack} of powerful radio emission.   If
radio-loud
and radio-quiet central engines are so similar, why should a line-of-sight
effect be so
important?  It must be that RQQs' central engines are located in a similar
nuclear environment, but one in which the observer is more likely to view the
central engine
through dusty, low- and high-ionization outflows.

We suggest that, for QSOs of the same Big Blue Bump luminosity, the same power
is available
to feed jets, that a `jet axis' exists in both RLQs
and RQQs, and that the observations can be accounted for by two main variables
-- one,
the inclination of the jet axis to the line-of-sight, and the other, the
inclination
of the jet axis to the plane of the host galaxy.  These possibilities are
illustrated
in Fig. 1.  The edges of the `ionization cone' within which the NLR can be
excited
are defined by the shadow of a dusty, inner torus.  In several well-observed,
but
low-luminosity cases, the cone is fairly symmetric about the jet axis (N4261,
Circinus:
Urry \& Padovani 1995).
In the radio-loud case the jet axis is, in some observations of FR\,II radio
galaxies,
perpendicular to the plane of the host galaxy (Ekers \& Simkin, and references
therein;
Heckman et al. 1985), or a dust
lane, or even parallel to the rotation axis of the host galaxy or extended
emission-line gas. In RQQs the inner regions may be symmetric about the jet
axis, but
the outer `torus' may warp to match the galaxy plane; a synchrotron
photon and high-energy particle beam may pound dense gas and dust in the inner
regions,
ablating Fe\,II-rich gas that shields the NLR.
The range of possible inclinations of the jet axis to the plane of the galaxy
could be
quite large, depending on the vertical thickness of dense material near the
nucleus.
This geometry would determine the relative numbers of RLQs and RQQs.  The
geometry
need not even be as simple as illustrated, for example, in the case of merging
galaxies.

Our proposed picture could explain differences in inclination dependence.  A
wider
range of viewing angles available for the inner parsec could explain the
jet--observer
inclination-dependence of axisymmetric emission regions in RLQs compared with
RQQs.
RLQs' `illumination cones' are well-defined by the dusty torus and
are free of the Galactic plane.  Still, grazing views of the torus can result
in
associated absorption and reddening.  Reddening can also result from a
low-latitude
view of the galactic plane, especially if the jet--cone axis is tilted towards
the
observer and the galactic plane.  Jet--observer inclination dependence for RQQs
is less
easily defined because galactic obscuration is important, and the `torus'
geometry may
be more complex.  Greater nuclear dust-covering may result in stronger Fe\,II
emission.
The inverse Fe\,II--[O\,III] relation --- increasing [O\,III] and
decreasing Fe\,II emission -- could result as the angle between the jet--cone
axis
and the galactic plane increases, illuminating more, distant, low-density NLR
gas.
This also explains the weaker Fe\,II and stronger NLR emission in RLQs.

The dustier, low-ionization absorption environment seen towards RQQs may not be
only
a jet--galactic plane inclination dependence.  It
has been commonly thought that powerful radio-loud AGN occur in elliptical
galaxies, and
the radio-quiet AGN occur in spirals (Hutchings et al. 1989).  This would tie
in nicely because
spiral galaxies are thought to be, more often, richer in gas and dust.  This
host-galaxy
dichotomy has also been suggested to relate to the nuclear and host galaxy mass
(Heckman 1983) or
angular momentum (Wilson \& Colbert 1995) hypotheses for the generation and
maintainance
of powerful radio jets, and it would be necessary to measure all three
parameters (mass,
angular momentum, and dusty, low-ionization environment) to disentangle these
hypotheses.
We note that elliptical galaxies can contain significant amounts of dust.
There are apparently RQQs in elliptical hosts (Disney et al. 1995), and
radio-loud sources in spiral galaxies, although tidal tails produced by mergers
that are
thought to fuel AGN, could be mistaken for spiral arms.  Are
RQQs ever found in ellipticals?  Hubble Space Telescope imaging may provide
the answer.

Why the lack of powerful jets in RQQs?
We suggest that this has something to do with the inner jet being within the
galactic
plane -- perhaps lack of collimation as a result of greater mass densities, or
perhaps
related to
orientation of the jet and galactic magnetic fields.  Radio cores may appear
weaker
as a result of absorption by highly-ionized nuclear plasma.  It may be a
problem that
light relativistic fluid jets are likely to propagate unimpeded through the
galactic
plane (Leahy, this workshop).

Tests of such a picture could be to investigate, by radio and
optical-ultraviolet
imaging, polarimetry, and spectroscopy, the relative orientation of some RQQs'
weak jets, possible illumination cones, and host galaxy orientation.  One could
look
for absorption in the nuclear, radio-core spectrum, and investigate its
possible
relation to absorption seen in the optical, ultraviolet and X-ray regions.

Having found significant differences between the spectra of RLQs and RQQs,
one could question our original assumption of the similarity of the central
engines of
RLQs and RQQs.  However, we might argue that the greatest differences we find
are
probably in gas thought to exist at least $\sim$ 1 pc from the central engine.

\end{document}